\documentclass[11pt]{article}

\usepackage[utf8]{inputenc}
\usepackage{tikz}      
\usepackage[margin=30mm]{geometry}
\usepackage{bm}
\usepackage{amsmath}
\usepackage{graphicx}
\usepackage{hyperref}
\usepackage{authblk}
\usepackage{pdfpages}

\title{Modelling of magnetic vortex microdisc dynamics under varying magnetic field in biological viscoelastic environments}
\author[1,2]{Andrea Vison\`a\thanks{corresponding author: andrea.visona96@gmail.com}}
\author[2]{Robert Morel}
\author[2]{H\'el\`ene Joisten}
\author[2]{Bernard Dieny}
\author[1]{Alice Nicolas}
\affil[1]{Univ. Grenoble Alpes, CNRS, CEA/LETI-Minatec, Grenoble INP, LTM, Grenoble F-38000, France }
\affil[2]{Univ. Grenoble Alpes, CEA, CNRS, Spintec, Grenoble F-38000, France}
\date{}
\usepackage[backend=biber,style=numeric-comp,sorting=none, maxbibnames=3, mincitenames=1, giveninits=true]{biblatex}
\begin{document}

\maketitle

\section*{Abstract}
\noindent{\small Magnetically driven microparticles provide a versatile platform for probing and manipulating biological systems, yet the physical framework governing their actuation in complex environments remains only partially explored. Within the field of cellular magneto-mechanical stimulation, vortex microdiscs have emerged as particularly promising candidates for developing novel therapeutic approaches. Here, we introduce a simplified two-dimensional model describing the magneto-mechanical response of such particles embedded in viscoelastic media under varying magnetic fields. Using a Maxwell description of the medium combined with simplified elasticity assumptions, we derive analytical expressions and support them with numerical simulations of particle motion under both oscillating and rotating magnetic fields. Our results show that rotating fields typically induce oscillatory dynamics and that the transition to asynchronous motion occurs at a critical frequency determined by viscosity and stiffness. The amplitude and phase of this motion is governed by the competition between magnetic and viscoelastic contributions, with particle motion being strongly impaired when the latter dominate. Energy-based considerations further demonstrate that, within the frequency range explored of few tens of Hertz, no heat is generated -distinguishing this approach from magnetic hyperthermia- while the elastic energy transferred to the surrounding medium is, in principle, sufficient to perturb major cellular processes. This work provides a simple framework to anticipate the first-order influence of rheological properties on magnetically driven microdisc dynamics, thereby enabling a better understanding of their impact in cells or extracellular materials and bridging the gap between experimental observations and theoretical modelling.}
\newpage

\section{Introduction}

In recent years, actuation of magnetic micro- and nano-objects using magnetic fields has emerged as a powerful tool for probing the mechanical properties of biological materials or for mechanically stimulating or altering cellular functions. These developments have been conducted both for improving our fundamental understanding of cell mechanobiology, in the form of magnetic tweezers \cite{sarkar_cancer_2013} and other micro-rheometers \cite{berret_local_2016}, and as proofs of concept for innovative therapeutic roads relying on the magneto-mechanical stimulation of cells \cite{dieny_magnetism_2025}, \cite{naud_cancer_2020}.  In both cases, the magnetic particles interact with a viscoelastic environment, the latter being either an inert material - such as a network of extracellular matrix proteins - or a responsive medium - such as cellular interior. Indeed, mechanical stimulations trigger cellular responses based on the fact that cells generate internal forces and adapt their physical properties in response to mechanical and biochemical stimuli to maintain homeostasis \cite{hoffman_cell_2009}. Common features of  magnetic particles used to mechanically stimulate cells or cellular environments are: a large shape anisotropy, in order to allow the magnetic actuation to be effectively translated into mechanical motion; and a large magnetic volume to exert forces up to the nanoNewton range. Such forces may be effective in deforming architectures of biological molecules, whose stiffness is often in the kiloPascal range, denaturing proteins or breaking filamentous proteins \cite{dobson_remote_2008}. Typical examples are micron-sized rods and discs \cite{naud_cancer_2020}. In particular, the disc-shaped particles have been extensively investigated to stimulate cells mechanically.  

\begin{figure}[h]
\centering
  \includegraphics[height=2.8cm]{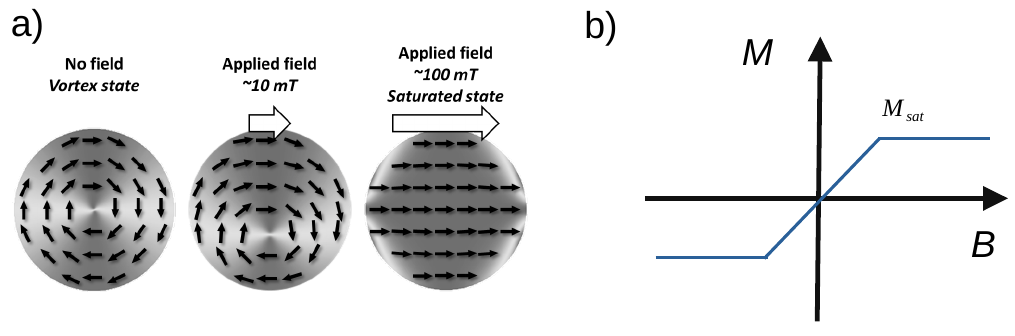}
    \caption{a) Illustration of the vortex micromagnetic behaviour in a disc-shaped particle: from vortex state (absence of external field) to saturation. b) Schematic representation of superparamagnetic-like behaviour of the vortex microdiscs. There is no remanent magnetisation in the absence of magnetic field. Particles made of permalloy with radius of the order of 1 $\mu$m and thickness below 100 nm saturate with field amplitude in the range 40-100 mT depending on their thickness \cite{leulmi_comparison_2013}.}
    \label{fig:fig1}
\end{figure}

For instance, in a pioneering work, Kim et al. \cite{kim_biofunctionalized_2010} demonstrated that magnetic actuation of micrometric disc-shaped particles at low frequency may trigger apoptosis with a success rate up to 90\% in a human glioblastoma model cell line. The particles were 60 nm thick, 1 $\mu$m diameter, Fe$_{20}$Ni$_{80}$ (permalloy) discs coated with 5 nm thick gold layers. Such particles, now commonly referred to as vortex microdiscs, have been used to mechanically stimulate cells, for instance to destroy cancer cells \cite{leulmi_triggering_2015}, activate neuronal signalling pathways \cite{gomez_elucidating_2023} or stimulate insulin secretion in pancreatic cells \cite{ponomareva_magnetic_2022}. The success of such particles comes from their magnetic properties. In absence of external magnetic field, the magnetisation consists in an in-plane closed loop yielding zero in-plane magnetisation and a very small out-of-plane magnetised core of dimension $\sim$5 nm at the centre of the microdisc. (Fig. \ref{fig:fig1}a) \cite{leulmi_comparison_2013}. Under a magnetic field, the particle gets magnetically polarised along the field yielding an in-plane shift of the vortex core in a direction perpendicular to the applied field direction (Fig. \ref{fig:fig1}). As the field is increased, the magnetic polarisation increases till vortex annihilation, leading to saturation at a field amplitude of typically few hundred mT. Due to its large shape anisotropy, the diameter of the disc being much larger than its thickness, the magnetisation tends to remain in the plane of the particle. This behaviour, characterised by zero magnetisation at zero field and gradual polarization under magnetic field, is referred to as superparamagnetic-like. Vortex microdiscs are therefore very promising for biomedical applications since the movement of the particles can be turned on and off effectively with magnetic fields and no remanent magnetisation is left when the field is turned off (Fig. \ref{fig:fig1}b), thus avoiding agglomeration of particles due to magnetostatic interactions if they are dispersed in solution \cite{leulmi_comparison_2013}.

In the frame of the aforementioned biological and biomedical applications, these particles are manipulated by low frequency rotating or oscillating fields. Such fields can be generated either by the motion of permanent magnets, such as the orbital motion of planar Halbach arrays \cite{halbach_design_1980} or of cylindrical magnets (for oscillating and rotating field respectively), or with coils and AC currents \cite{kim_biofunctionalized_2010}. The field magnitudes used are in the order of tens to hundreds of mT, depending on the magnetic material and the geometry of the particles. The frequencies employed range between 0.5 to 20 of Hz \cite{naud_cancer_2020}. 

Despite many experimental examples of the applicability of such technological approach, little has been investigated about the physical interaction of the particles with both the surrounding medium and the magnetic field. Only two studies have proposed a physical description of the mechanical stresses exerted by the magnetised vortex discs, and compare it with typical order of magnitude of biomolecular stresses. For instance Kim et al. \cite{kim_biofunctionalized_2010} calculated the mechanical torque exerted by a disc attached to the cell membrane. The problem was solved in the quasi-static regime, corresponding to the situation where the disc movement is halted by the elastic torque opposed by the membrane. In this study the dependence of the magnetisation to the magnetic field was neglected. In a second study, Leulmi et al \cite{leulmi_comparison_2013} proposed an in-depth calculation of the magnetic torque exerted by the particles with a more complete description of the magnetic problem. However, the reaction of the outer medium was not modelled, as the particle was assumed to be immobilized  on the substrate.

In this study, we address the modelling of the vortex microdiscs movement in a viscoelastic environment when exposed to oscillating or rotating fields. We aim to couple the magnetic and viscoelastic problem in a more comprehensive way as was done before. For instance, our goal is to provide a dynamic analysis of the movement of the vortex particle as a function of the viscoelastic properties of the outer medium. We propose here a first order description of the magnetic problem and of the elastic resistance of the material in which the particle is embedded. Our objective is to provide a simple framework for anticipating the impact of the rheological properties of the microdisc environment on its movement, thereby enabling a better understanding of its magnetic actuation within cells or extracellular materials. For the sake of simplicity, we limit our analysis to a 2D description. By doing so, we constrain the particle rotation around the axis perpendicular to the rotation plane of the field. This amounts to imposing anisotropic mechanical properties on the material in which the microdisc is embedded, which prevents it from aligning with the rotation plane of the field. In this assumption framework, an important result of this study is that the magnetic torque induces a mechanical torque whose amplitude is given by the orientation of the magnetisation relative to the plane of the particle. This leads us to show that the elastic resistance of the outer medium can either impair the rotation of the particle or favour it when its relaxation also relaxes the magnetic problem. Finally we show that viscosity, by imposing a lag in the movement of the particle relative to the magnetic field, can in some conditions convert the influence of a rotating field into an oscillating motion. Eventually, we compare the orders of magnitude of the energies at play to common biomolecular events to conclude on the ability of the movement of the particles to mechanically influence biological processes. 

\section{Theory}
We consider magnetic particles made of ferromagnetic material with highly anisotropic geometric specifications, with radius $R$ and thickness $h$. In the absence of field, when $R \gg h$, the magnetisation organizes as a vortex (Fig. \ref{fig:fig1}a) \cite{guslienko_magnetic_2008}. Its magnitude in every unit volume is the spontaneous magnetisation, also called the saturation magnetisation $M_{sat}$. When exposed to an external field of sufficient amplitude, the magnetisation vectors align (Fig. \ref{fig:fig1}a), and the particle experiences a magnetic moment whose amplitude is:

\begin{equation}
    m_{sat} = M_{sat}V 
\end{equation}
where $V$ is the volume of the disc.

\begin{figure}[h]
\centering
  \includegraphics[height=8cm]{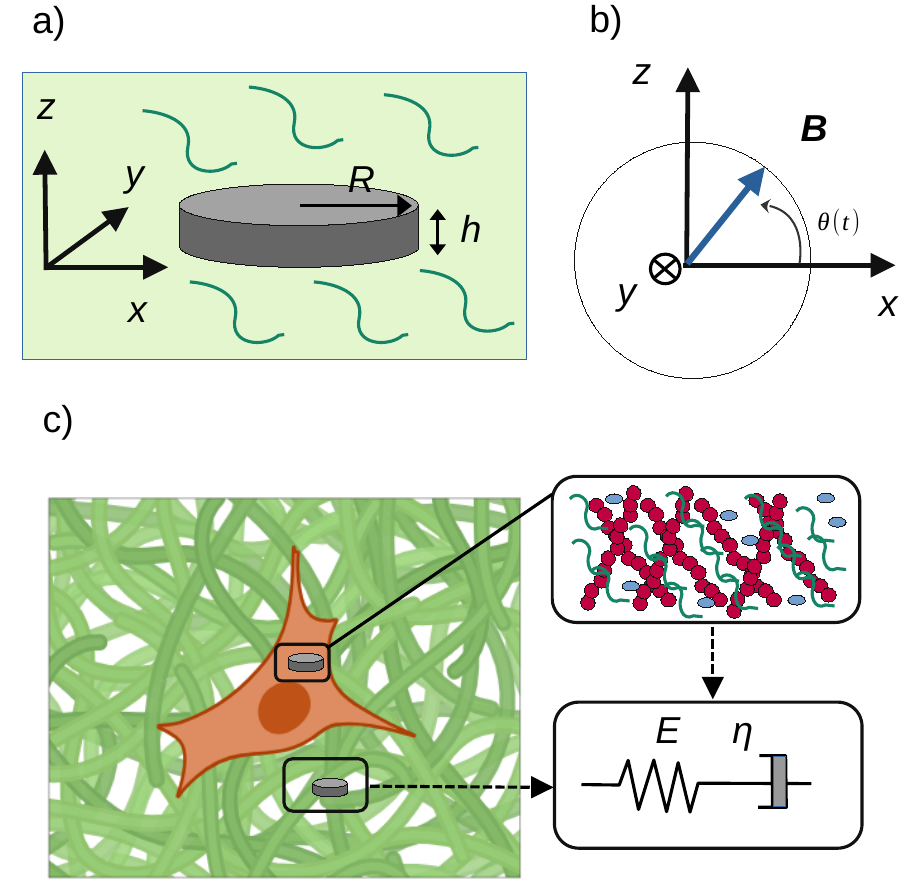}
    \caption{a) Microdisc embedded into a 3D viscoelastic material. b) 2D representation of an external field, rotating or oscillating in the ($x,z$) plane. The field orientation at time $t$ is given by $\theta(t)$. The field is invariant in the $y$ direction. c) The particle is trapped into a material, either extracellular or intracellular, approximated by a Maxwell viscoelastic material. Its rheological properties are equivalent to a spring and a dashpot in series.}
    \label{fig:fig2}
\end{figure}

Here we consider a particle that is embedded in a viscoelastic material as depicted in Fig. \ref{fig:fig2}. We assume that in its resting position, the plane of the particle lies in the $(x,y)$ plane of the reference frame. In the following, we focus on the effect of rotating or oscillating magnetic fields. No field gradients are considered, which would result in the translational movement of the particle. The magnetic field, denoted $\mathbf{B}$, is modelled as a constant amplitude field which may rotate or oscillate at a certain frequency, $f$, in the $(x,z)$ plane (Fig. \ref{fig:fig2}b). 
We assume that the field is invariant in $y$ direction. Its analytical expression is: 

\begin{equation}
    \mathbf{B} = B_0 
        \begin{pmatrix}
            \cos\theta(t)\\
            0\\
            \sin\theta(t)
        \end{pmatrix}
        \label{eq:field_mat}
\end{equation}
where $\theta(t)$ is the angle between the field vector and the $x$ axis. In case of an oscillating field, the expression of $\theta(t)$ is $\theta(t)=\theta_0\sin\omega t $ with $\omega=2\pi f$ and $\theta_0$ the amplitude of the oscillation. For a rotating field instead, $\theta(t)=\omega t$. 
This simplified field model captures the behaviour of most experimental devices (reviewed in Naud et al \cite{naud_cancer_2020}), where the magnetic field is essentially uniform along the $y$-axis when measurements are taken sufficiently far from the edges, where boundary effects are dominant. Out of these regions, the field rotates or oscillates within a specific plane. Once an external field is applied, the magnetised particle experiences a magnetic torque which tends to align the plane of the particle to the direction of the field. Such torque is counteracted by viscous and elastic resistances coming from the external medium. In the following, we propose a physical description of the interaction of the magnetic field with the microdisc, and of the viscoelastic reaction of the surrounding medium. We then analyse the effect of these viscoelastic properties on the motion of the microdisc driven by oscillating or rotating magnetic fields.

\subsection{Modelling vortex microdiscs response to an external field}

The magnetic configuration in absence of any external field is the one of a closed vortex flux (Fig. \ref{fig:fig1}a). Such magnetic texture mainly arises from two competing effects, the self-magnetisation of the material and the appearance of an internal, demagnetising field. The first effect, resulting in the so-called exchange energy $E_{ex}$ \cite{coey_magnetism_2010}, arises from quantum mechanical interactions between neighbouring electron spins (Coulomb repulsion of the electrons and Pauli exclusion principle). In ferromagnetic materials, it favours the  parallel alignment of the spins and leads to spontaneous magnetisation. The second effect, giving rise to the demagnetisation energy $E_{d}$ and also referred to as shape anisotropy, tends to maintain the spins oriented in the plane of the particle.  Phenomenologically, the demagnetising field can be described as the consequence of magnetic charges that appear at the surface of the magnetic material when the spins are out-of-plane. This energy term favours minimal demagnetising field, which corresponds to having the magnetic charges as far as possible. In principle, the crystalline structure of the magnetic material also influences the magnetic configuration. Here we neglect the effect of the magnetocrystalline anisotropy which tends to align the magnetisation with some specific orientation of the crystalline structure. For example, in the case of permalloy, a magnetic material often employed in the design of vortex particles, its contribution vanishes due to the opposite contributions of iron and nickel atoms.

When an external magnetic field is applied, the Zeeman energy, $E_{Z}$, which accounts for the tendency of the magnetisation to align with the magnetic field, has to be included. Thus the total magnetic energy reads:

\begin{equation}
    E_{tot} = E_{ex} +  E_{d} + E_{Z}
    \label{eq:mag_energy}
\end{equation}

Depending on the amplitude of the external field, the vortex core is progressively displaced until it is completely ejected from the disc, and the particle becomes fully magnetised (Fig. \ref{fig:fig1}a). For such particles, magnetic fields in the mT range are enough to saturate the magnetisation \cite{kim_biofunctionalized_2010,leulmi_triggering_2015,cheng_rotating_2016}.

\begin{figure}[h]
\centering
  \includegraphics[height=3.4cm]{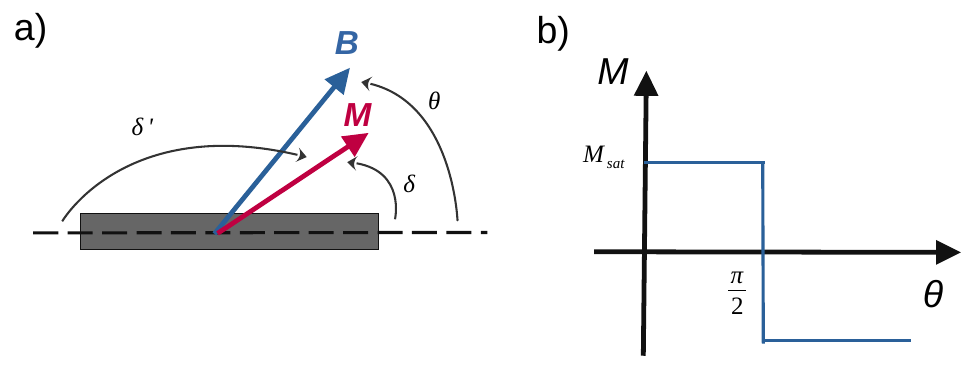}
    \caption{a) The Stoner-Wohlfarth model determines the orientation of the magnetisation $\bf{M}$ relative to the plane of the particle (labelled by the angles $\delta$ or $\delta' = \pi - \delta$) as a function of the orientation of the field $\bf{B}$ (angle $\theta$). b) In our model, the amplitude of the field is assumed constant while its orientation varies in time. We approximate the magnetic hysteresis curve to first order: the magnetisation of the disc flips from one saturated state to the other when the field is perpendicular to the plane of the disc.}
    \label{fig:fig3}
\end{figure}

For the scope of this paper, we are not interested in the dynamics of the vortex core, and we treat the magnetic problem as if the particle is always saturated (Fig. \ref{fig:fig3}). In doing so we use the macrospin model which is often employed to describe the magnetisation dynamics of single-domain magnetic particles \cite{coey_magnetism_2010} in the framework of the Stoner-Wohlfarth model (Fig. \ref{fig:fig3}a). The magnetisation field is then reduced to a single, uniform vector $\bf M$. This approximation is governed by the Landau–Lifshitz–Gilbert equation. Since the frequency of the external field is much lower than the Larmor frequency (the order of magnitude of the gyromagnetic ratio is GHz/T$^{-1}$ in ferromagnetic materials such as permalloy \cite{noginova_ferromagnetic_2021}), we neglect the dynamics of the magnetisation. We calculate its orientation as a result of the equilibrium between Zeeman and shape anisotropy energy terms, the contribution of the exchange energy being negligible \cite{yin_magnetocrystalline_2006}: 

\begin{equation}
     E_{tot} = E_{d}(\delta) + E_{Z}(\delta)
    \label{eq:mag_energy_simp}
\end{equation}
Eq. \ref{eq:mag_energy_simp} depends on a single angle, $\delta$, the angle between the plane of the particle and the magnetisation vector (Fig. \ref{fig:fig3}a). In this model, the magnetisation of the microdisc is assumed to be saturated at all times. The variation of the magnetisation vector relative to the orientation of the applied field is therefore approximated as a step function, as reported in Fig. \ref{fig:fig3}b. This implies that the magnetisation flips in orientation once the angle $\theta$ becomes larger than $\pi/2$. 

An analytical expression of the demagnetising energy can be derived by approximating the thin disc to an oblate ellipsoid whose principal axis in the $(x,y)$ plane are of identical length, approximated to be the radius $R$ of the microdisc, and the out-of-plane axis is approximated to be of length $h$, the thickness of the disc (see Supplementary information):

\begin{equation}
E_d = \frac{1}{2}\mu_{0}\frac{m_{sat}^2}{V}\left( N_R \cos^2\delta + N_h \sin^2\delta\right)\label{eq:Edemag}
\end{equation}

In this equation, $N_R$ and $N_h$ are the two demagnetising coefficients that depend on the aspect ratio $R/h$. For a true oblate geometry, they are related by the condition $N_R = \frac{1 - N_h}{2}$. In our case, since the geometry departs from an oblate, this relation does not hold. However, we could show using a micromagnetic simulation that Eq. \ref{eq:Edemag} is suitable to describe the demagnetizing energy but without the aforementioned relationship between $N_R$ and $N_h$ (article in preparation). The numerical values obtained for permalloy microdiscs are gathered in Table \ref{tab:param}.

In the presence of a magnetic field, the magnetisation $\bf{M}$ is attracted toward the field. The Zeeman energy, that describes this attraction, writes:

\begin{equation}
    E_Z = - \mathbf{m}\cdot\mathbf{B} = -m_{sat}B_0\cos(\theta-\delta) \label{eq:Ez}
\end{equation}
with $\bf m$ the magnetic moment associated to $\bf M$: ${\bf m} = {\bf M}V$.

\subsection{Model for the mechanical response of the medium to the particle motion}

We assume that the particle is embedded in a Maxwell-like material (Fig. \ref{fig:fig2}c). This modelling is the most simple approach often used to describe the rheological properties of either extracellular matrices or the cell bodies \cite{bonfanti_fractional_2020}. In this study, we limit our description to a 2D framework so to be able to propose analytical analysis. Since the particle is not invariant in a specific direction, this assumption amounts to constraining its rotation to the $y$-axis, which is the axis of rotation of the magnetic field. Under this assumption, the material has highly anisotropic mechanical properties, but stress relaxation remains  characterised by one single relaxation time:

\begin{equation}
\tau = \frac{\eta}{E}\label{eq:relaxT}
\end{equation}
where $\eta$ is the viscosity and $E$ the elastic modulus of the material along the deformable direction. The equivalent circuit associated to the Maxwell model is a spring in series with a dashpot, submitted to a unidirectional force $F$ (Fig. \ref{fig:fig1}c). The displacement $x$ of the particle is then described by Eq. \ref{eq:max_lin}: 

\begin{equation}
    \frac{dx}{dt} = \frac{1}{k}\frac{dF}{dt} + \frac{F}{\xi} \label{eq:max_lin}
\end{equation}
where $k$ and $\xi$ are the spring elastic constant and the dashpot viscous constant, respectively. These are proportional respectively to the elastic modulus $E$ and to the viscosity $\eta$ of the material by a geometrical factor depending on the shape of the particle and the direction of the force relative to the orientation of the particle. This equation can be transposed to a rotating geometry when the particle is submitted to a torque, $\Gamma$ \cite{wilhelm_local_2002}. In this case, the Maxwell model  describes the angular velocity  $\frac{d\alpha}{dt}$  of the particle in the Maxwell-like material (Fig. \ref{fig:fig4}).

\begin{equation}
    \frac{d\alpha}{dt} = \frac{1}{\gamma}\frac{d\Gamma}{dt} + \frac{\Gamma}{\nu}\label{eq:max_rot}
\end{equation}
where $\gamma$ and $\nu$ are the torsional and viscous constants.

\subsubsection{Modelling the viscous resistance of the medium to the particle motion}

\begin{figure}[h]
\centering
  \includegraphics[height=6cm]{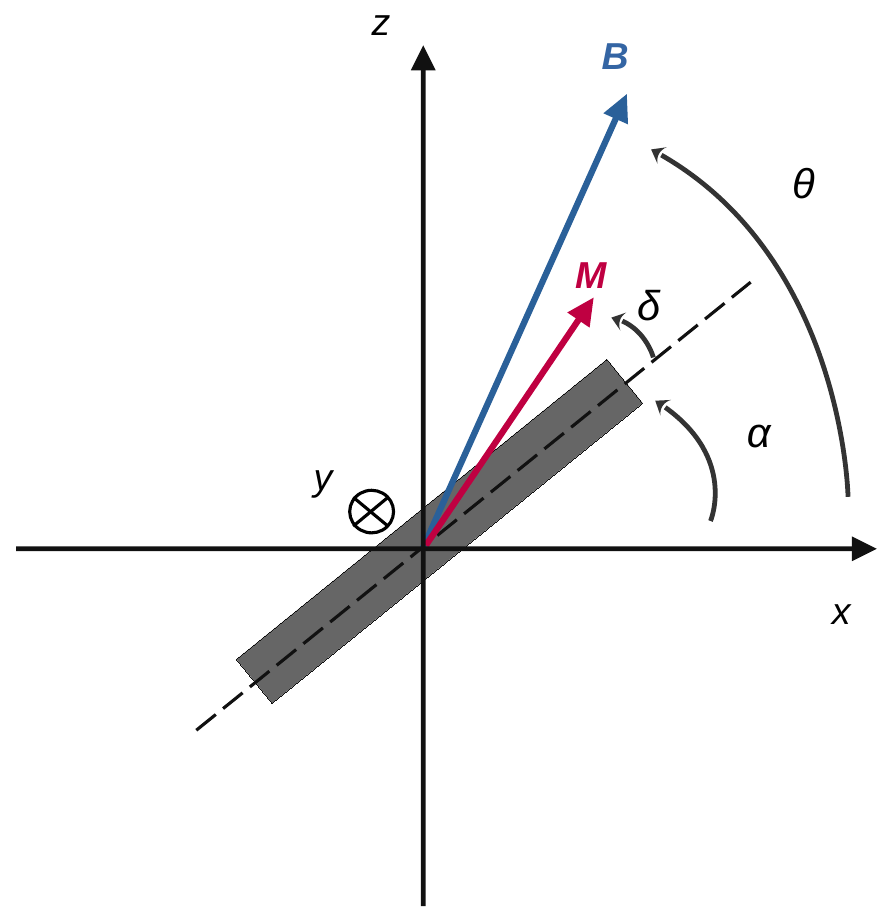}
    \caption{Geometrical parameters of the model when the particle can rotate.}
    \label{fig:fig4}
\end{figure}

As a consequence of the magnetic actuation, the microdisc may rotate from its resting position around the $y$-axis. This introduces the angle $\alpha$ which measures the rotation of the particle. The dynamics of $\alpha$ is linked to the viscous and elastic resistance of the external material (Fig. \ref{fig:fig4}). 

We first calculate the viscous torque that opposes to the field-induced rotation. Since the particle is rotating around one of its symmetry axis that is parallel to its plane, in the regime of laminar flow, the viscous torque scales with the radius of the particle and not with its thickness. Following Ref. \cite{guyon_book} and dropping the shape-dependent numerical factors, its expression is:
\begin{equation}
\boldsymbol{\Gamma} = \nu \dot{\alpha} \simeq - \eta R^3 \dot{\alpha}\, {\bf \hat{y}} 
\label{eq:Gvis}
\end{equation}
where $\dot{\alpha}=\frac{d\alpha}{dt}$ is the angular velocity of the particle, and ${\bf \hat{y}}$ the unit vector along the $y$-axis, which is the axis of rotation (Fig. \ref{fig:fig4}). A similar relation between $\nu$ and $\eta$ has been derived for a cylinder in \cite{doi_theory_1986} and used in \cite{wilhelm_local_2002}. The viscous energy loss then writes:
\begin{equation}
    E_{vis} \simeq \eta R^3 \int \dot{\alpha} d\alpha 
    \label{eq:Evis}
\end{equation}

\subsubsection{Elastic resistance of the medium to a small angle deformation}

We now model the elastic resistance the material opposes to the rotation of the particle.
In principle, the relation  between the local force, coming from the magnetic actuation and the deformation of the elastic material can be calculated by solving the equilibrium equation of elasticity \cite{landau_theory_2009}. A mathematical framework was proposed by Chadwick et al. \cite{chadwick_torsional_1969} to calculate the deformation field resulting from torsional oscillations of rigid particles of arbitrary shape embedded into an infinite elastic solid. So in principle, the exact energy cost induced by the rotation of the microdiscs can be calculated. But no analytical solution is available as soon as the shape of the particle departs from a sphere \cite{chadwick_oscillations_1966}. For our analysis, we made the choice to approximate the elastic resistance of the material to its very first order, so that we could keep an analytical expression for it. Nevertheless in the following, we expose the limitations of our approximations so that it is possible to envision their limits.

Our first approximation is to consider that the rotation of the particle is limited to small angles. This approximation will be challenged in the next sections. If the particle is rotated around one of its diameter, at small angle, the main resistance comes from the compression (and dilatation) of the material perpendicular to the flat planes of the particle. Shear stresses are thus neglected. This approximation is justified by the shape anisotropy of the microdiscs: the thickness of the particle is small compared to its diameter so that when the rotation is limited to small angles, the force component is mainly normal. Our second approximation is to neglect the contribution of the long range propagation of the elastic deformation. The long range propagation of the deformation indeed results in elastic energy being stored at distance of the moving particle, and potentially released toward the particle when the stress in the vicinity of the particle decreases \cite{chadwick_oscillations_1966}.
This leads us to model the elastic response of the material with a Hook's law:
\begin{equation}
f_n \simeq k u_n
\label{eq:hook}
\end{equation}
with $f_n$ the force per unit surface exerted by the microdisc normal to its plane, $u_n$ the normal displacement of the material in the vicinity of the particle and $k$ the elastic constant of the material, a function of the elastic modulus $E$, the Poisson ratio of the material and the radius $R$ of the particle (the thickness cannot contribute in this crude approach as no shear stress are considered). Dimensional analysis suggests that $k \sim E/R$.

In the framework we propose, the rotation of the particle induces a small normal displacement of the material, $u_n = r \alpha$, with $r$ the distance to the axis of rotation.  From Eq. \ref{eq:hook}, we therefore conclude that the elastic energy generated by the motion of the particle has the following scaling with the rotation angle:
\begin{equation}
E_{el} \simeq \int_S dS \int f_n r d\alpha \simeq \frac{1}{2}\gamma \alpha^2
\label{eq:Eel}
\end{equation} 
where $\gamma = \frac{\pi R^4}{4}k$ is the approximate torsional constant.

\subsection{Equation governing the motion of a saturated vortex disc in a viscoelastic medium}

Coupling the magnetic (Eqs \ref{eq:Edemag} and \ref{eq:Ez}) and mechanical problems (Eqs \ref{eq:Evis} and \ref{eq:Eel} ), the total energy of the system writes:

\begin{eqnarray}
    E_{tot} &=& E_{kin}( \alpha) + E_{mag}(\delta, \theta-\alpha) + E_{mech}(\alpha) \label{eq:Edynam}\\
     &=& \frac{1}{2}I\dot{\alpha}^2 + \frac{1}{2}\mu_{0}\frac{m_{sat}^2}{V} \left( N_R \cos^2\delta + N_h \sin^2\delta\right) + \\ &-& m_{sat}B\cos( \theta-\alpha - \delta ) + \frac{1}{2} \gamma \alpha^2 + \nu \int \dot{\alpha} d\alpha \nonumber
\end{eqnarray}
where $I \simeq \frac{m\pi R^2}{4}$ is the moment of inertia. Since the mass of a magnetic microparticle is of the order of $10^{-15}$ kg, the kinetic energy term ($E_{kin}$) is negligible compared to the other energetic terms (see Table \ref{tab:param}).  

Eq. \ref{eq:Edynam} has two independent variables, $\delta$ and $\alpha$. The values they assume when the magnetic field is oriented with an angle $\theta$ relative to the lab frame are found by minimising the total energy respective to these two quantities:
   
\begin{eqnarray}
      K_D\sin2\delta &=& m_{sat}B_0\sin(\theta-\alpha-\delta)\label{eq:mag}\\
      \gamma\alpha + \nu\dot{\alpha} &=& m_{sat}B_0\sin(\theta-\alpha-\delta) \label{eq:mec}
\end{eqnarray}
with $K_D = \frac{\mu_0}{2} \frac{m_{sat}^2}{V} (N_h-N_R)$ characterizing the energetic cost of the magnetic anisotropy.

\section{Results}

\subsection{Tuning the numerical parameters of the model}
Eqs. \ref{eq:mag}--\ref{eq:mec} were solved for a range of parameters relevant for applications involving biological cells or tissues, in which such anisotropic microparticles have been used. The list of all the parameters used in the model and their respective numerical values are reported in Table \ref{tab:param}.

\begin{table}[h]
\small
\centering
  \caption{\ Summary of the parameters used in the model, their symbol and numerical values}
  \label{tab:param}
  \begin{tabular*}{0.6\textwidth}{@{\extracolsep{\fill}}lll}
    \hline
    Parameter & Symbol & Numerical value\\
    \hline
        Disc radius & $R$ & 0.65 $\mu$m \\
         Disc thickness & $h$  & 60 nm \\
         Disc magnetic moment & $m_{sat}$ & $6.37\times 10^{-14}\text{A.m}^2$ \\
         Demagnetising factors & $N_R, N_h$ & 0.0517 and  0.7076 \\
         Magnetic field amplitude & $B_0$ & 100 mT \\
         Field frequency & $f$ & [1-100] Hz\\
         Anisotropy constant & $K_d$ & 2.1$\times10^{-14}$ J \\
         Young's modulus & $E$ & [0.05 - 10] kPa\\
         Viscosity & $\eta$ & [10-1000] Pa.s \\
                
    \hline
  \end{tabular*}
\end{table}

The sizes of the microdiscs were inspired by Refs. \cite{kim_biofunctionalized_2010,leulmi_triggering_2015, cheng_rotating_2016, ponomareva_magnetic_2022} that employ such particles to mechanically stimulate cells, as well as the range of frequency. The amplitude of the field was chosen so that the magnetisation is saturated. Furthermore, it was shown that ferromagnetic particles capable of vorticity, such as those made of permalloy, reach saturation with magnetic fields below 100 mT \cite{leulmi_comparison_2013}. The orders of magnitude of the mechanical properties of the viscoelastic medium were chosen based on the rheological studies of intracellular compartments or of common biomaterials such as Matrigel, Collagen or Hyaluronic scaffolds. For instance, common values of the Young's modulus of Matrigel span between few tens of Pa to 2 kPa \cite{soofi_elastic_2009}, while its viscosity is around 50 Pa.s \cite{de_stefano_bioprinting_2021}.  Collagen and Hyaluronic scaffolds also have stiffness that ranges between hundreds of Pa to several kPa, depending on their crosslinking and concentration \cite{su_adipogenesis_2022}.

As far as cell rheology is concerned, we assume that the particle is trapped in the cellular cytoskeleton or immersed in the cytoplasm. Given that the literature provides spread values with very different orders of magnitude of elastic modulus and viscosity for actin cortex \cite{hoffman_cell_2009,lim_mechanical_2006,alibert_are_2017}, we decided to set a range of elastic modulus that spans between 1 kPa to 10 kPa to account for different crosslinking states, interplay with intermediate filaments and microtubules \cite{grady_cell_2016,rotsch_drug-induced_2000,guo_role_2013}. Concerning the cytoplasm, we chose lower values for the elastic modulus, between 0.05 kPa and 0.5 kPa \cite{berret_local_2016}. The same approach was used to set a range of possible values of viscosity, between 10 and 1000 Pa$\cdot$s. This range aims at taking into consideration different polymerisation states of the cytoskeleton. For instance, we associated higher viscosity to well organised networks \cite{ketene_actin_2012}.

The final ranges of Young's moduli and viscosities are reported in Table \ref{tab:param}.

\subsection{Shape anisotropy term imposes the angle between the easy plane and the magnetisation, $\delta$, to be small}

A first limit case of an immobilised particle is modelled to get information on the magnitude of the angle $\delta$ between the magnetisation $\bf{M}$ and the easy plane of the particle (Fig. \ref{fig:fig3}a). In this case, the solution of the problem is provided by Eq. \ref{eq:mag}  where $\alpha$ is set to 0:

\begin{equation}
    K_d\sin2\delta - m_{sat}B_0\sin({\theta-\delta}) = 0
    \label{eq:SW}
\end{equation}

\begin{figure}[h]
\centering
  \includegraphics[height=5.5cm]{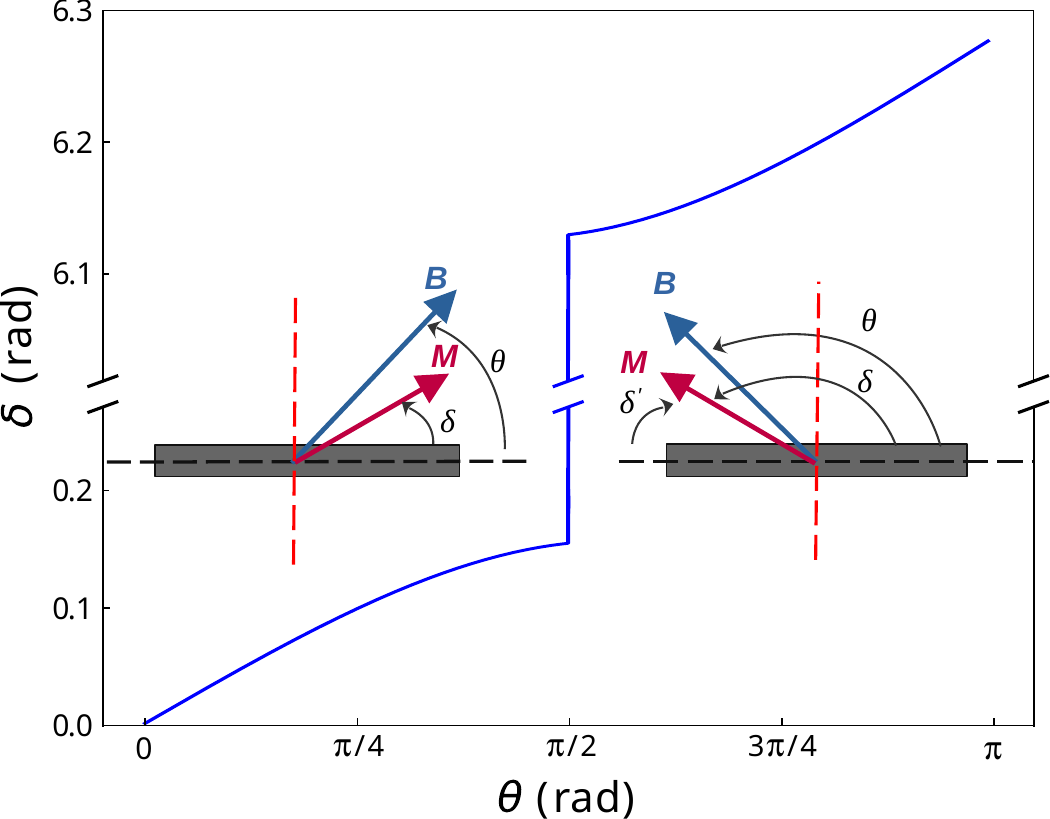}
    \caption{Evolution of the angle $\delta$ between the anisotropy plane and the field as function of the field angle, $\theta$, for an immobilised particle (Stoner-Wohlfarth's model).}
    \label{fig:stoner}
\end{figure}

The equation is solved for $\theta$ between 0 and $\pi$. When $\theta$ exceeds $\pi/2$, the magnetisation flips in the symmetric direction relative to the $(y,z)$ plane. Nonetheless, the equation is not altered by this flip as demonstrated in the following. For values of $\theta \in[0, \pi/2]$, the equation describing the equilibrium state is Eq. \ref{eq:SW}. When $\theta \in [\pi/2, \pi]$, the magnetisation at equilibrium is obtained with a newly defined angle $\delta'=\delta-\pi$ (mind the negative sign of $\delta'$) (Figs \ref{fig:fig3}a and \ref{fig:stoner}). $\delta'$ is the angle that governs the physics of the system once the magnetisation has flipped. Eq. \ref{eq:SW} then becomes:

\begin{align}
      -K_d\sin(2\delta') - m_{sat}B_0\sin(\pi-\theta+\delta') &= \nonumber\\
      -K_d\sin2\delta + m_{sat}B_0\sin(\theta-\delta) &= 0 \label{eq:proof_flip} 
\end{align}

Expressed in terms of $\delta$,  Eq. \ref{eq:proof_flip} therefore remains identical to Eq. \ref{eq:SW}.

Eq. \ref{eq:SW} was solved numerically with an in-house python code for the numerical values reported in Table \ref{tab:param}. The values of $\delta$ are shown in Fig. \ref{fig:stoner}.
We therefore conclude that for an immobilized particle, either $\delta$ or $\delta'$ remains small while $\theta$ is varied. An immobilised particle corresponds to the limit case where $\mathbf{M}$ deviates the most from the particle plane. Therefore we assume that the conditions $|\delta|\ll1$ or $|\delta'|\ll1$ are met also when the particle is free to rotate.

\subsection{Adimensional parameters governing the movement}

Having demonstrated that the magnetisation only slightly deviates from the easy plane of the particle, Eqs. \ref{eq:mag} and \ref{eq:mec} can be simplified by expanding them to first order in $\delta \ll 1$:

\begin{eqnarray}
    \delta \simeq \frac{\sin(\theta-\alpha)}{b +cos(\theta- \alpha)}\nonumber\\
     n\dot{\alpha} + g\alpha - b\delta \simeq 0 \label{eq:delta}
\end{eqnarray}
where we have introduced reduced variables which compare the contributions of mechanical and shape anisotropy torques that all limit the reorientation of the magnetisation, to the Zeeman torque that tends to align the magnetisation to the magnetic field.

\begin{equation}\label{eq:reduced}
 n = \frac{\nu}{m_{sat}B_0}  \quad \textrm{,} \quad g = \frac{\gamma}{m_{sat}B_0} \quad \textrm{and} \quad b = \frac{2K_d}{m_{sat}B_0}
\end{equation}

The numerical values of these parameters are reported in Table \ref{tab:param2}. 

\begin{table}[h]
\small
\centering
  \caption{\ Summary of the reduced parameters used in the model, their symbol and numerical values}
  \label{tab:param2}
  \begin{tabular*}{0.6\textwidth}{@{\extracolsep{\fill}}lll}
    \hline
    Parameter & Symbol & Numerical value\\
    \hline
      Reduced magnetic coefficient & $b$ & 6.5 \\
         Reduced elastic coefficient& $g$ & [$2\times10^{-4}-2$] \\
         Reduced viscosity coefficient & $n$ & [$4\times10^{-5}-0.4$] s \\
    \hline
  \end{tabular*}
\end{table}

\subsection{Microdisc motion subjected to an oscillating magnetic field}

We first focus on the motion of a microdisc submitted to an oscillating magnetic field:
\begin{equation}
\theta(t)=\theta_0\sin(\omega t)
\label{eq:theta_oscil}
\end{equation}
We limit the study of the motion of the particle to values of $\theta_0 < \pi/2$. Magnetisation flipping events will be addressed later on, in the rotating field section. In the limit where $\theta - \alpha \ll 1$, which is expected either when the field oscillates at small angles ($\theta_0\ll1$) or when the torques that oppose to the Zeeman torque are small enough, Eq. \ref{eq:delta} can be solved analytically. Considering the initial condition $\alpha(0) = 0$, we find:

\begin{equation}
    \alpha(t) =  \theta_0\frac{C\omega}{A^2 +\omega^2 }\left( e^{-A t}+\frac{A}{\omega}\sin(\omega t) - \cos(\omega t)\right)
    \label{eq:alpha_oscil}
\end{equation}
with
\begin{align}
A &= \frac{g}{n} + \frac{b}{n (b+1)}\label{eq:A}\\
C& = \frac{b}{n(b+1)}\label{eq:C}
\end{align}
Note that the first term in $A$ is the inverse of the Maxwell time (Eq. \ref{eq:relaxT}).
Before drawing any conclusion, the reliability of the analytical solution Eq. \ref{eq:alpha_oscil} is tested against the numerical solution of Eq. \ref{eq:delta}. The numerical problem is solved with an in-house python code making use of \verb|solve_ivp| function of SciPy module \cite{scypy}.

\begin{figure}[h]
\centering
  \includegraphics[height=4.5cm]{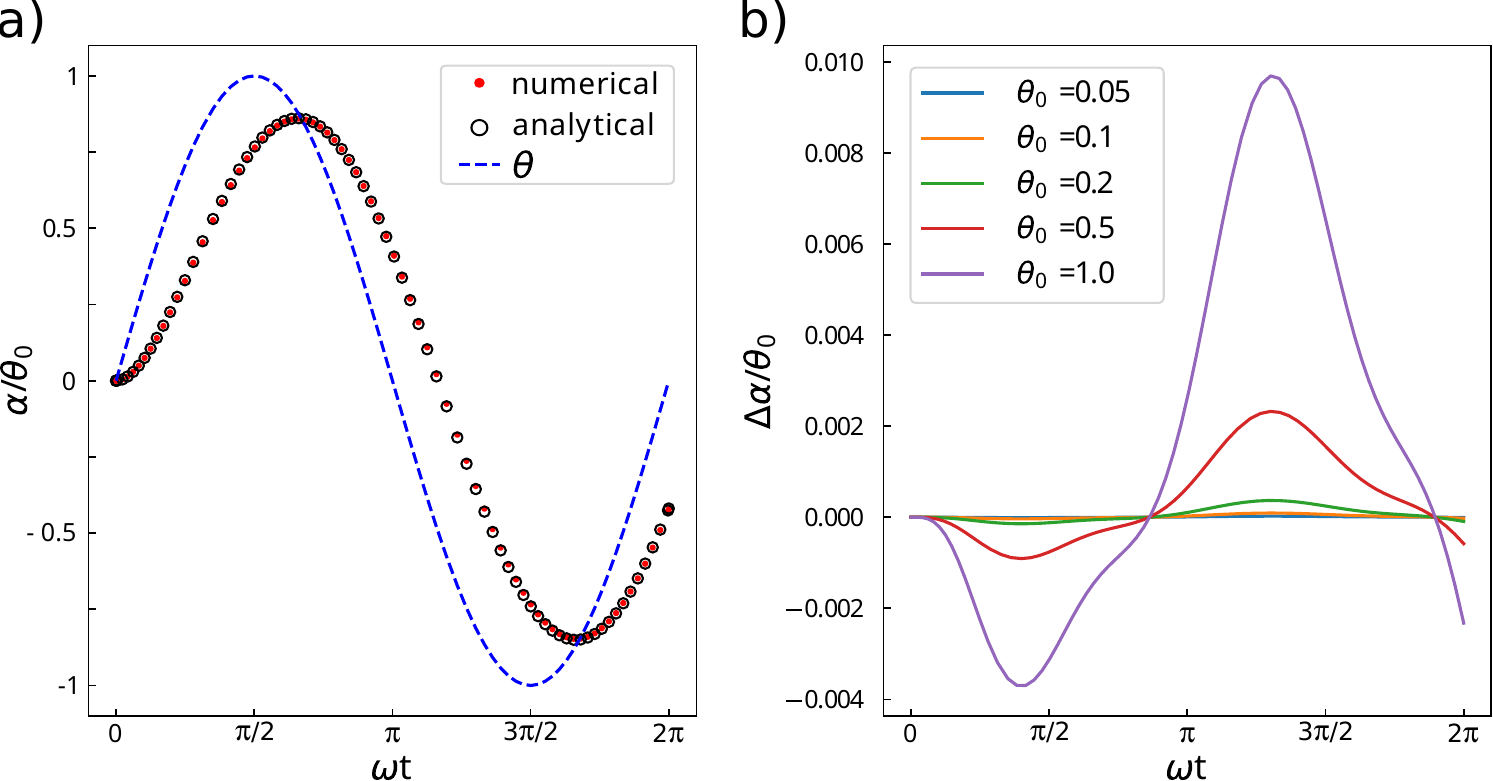}
    \caption{a) Comparison of analytical ($\circ$) and numerical ({\color{red}$\bullet$}) solutions for $\theta_0 = 1$ rad, calculated for intermediate values of $g$ and $n$ ($g = 2\times10^{-2}$, $n = 4\times10^{-2} $s). $\theta$ is shown to assess the assumption $\theta - \alpha \ll 1$. b) Error between the numerical and analytical solution for different values of $\theta_0$. $\Delta\alpha$ is calculated as the difference between the numerical solution and the analytical one provided by Eq. \ref{eq:alpha_oscil}.}
    \label{fig:comparison_num_analy}
\end{figure}

Fig. \ref{fig:comparison_num_analy} shows that the analytical solution remains valid with a good accuracy beyond the approximation $\theta_0\ll1$. The effectiveness of Eq. \ref{eq:alpha_oscil} in describing particle motion for oscillations of significant amplitude justifies the use of the analytical approach to deduce asymptotic motion behaviours as a function of solicitation frequency or rheological parameters of the particle's environment. Reported trends should be accurate to within a few percent, even for the largest oscillations (Fig. \ref{fig:comparison_num_analy}b).

Eq. \ref{eq:alpha_oscil} shows that $\alpha(t)$ reaches a stationary regime after a characteristic time $1/A$. This characteristic time comprises two contributions, that act in parallel: the relaxation of the elastic stress in the viscoelastic material following particle's movement, and the friction of the surrounding material that slows down the particle rotation induced by the magnetic torques (Eq. \ref{eq:A}). Once the steady state is reached, the oscillation of the microdisc forced by the magnetic field is a phase-shifted oscillation at the forcing frequency, whose amplitude decreases when the frequency is increased (Eqs \ref{eq:oscil_stat}--\ref{eq:oscil_phase}):
\begin{equation}
    \alpha(t) = \alpha_0 \sin(\omega t -\Phi) \label{eq:oscil_stat}
\end{equation}
with:

\begin{align}
&\alpha_0 = \theta_0\frac{C}{\sqrt{A^2 +\omega^2 }} =\theta_0\frac{\frac{ b }{(b+1)}}{\sqrt{(g + \frac{b}{(b+1)})^2+n^2 \omega^2}} \label{eq:oscil_amplitude}\\
&\tan(\Phi) = \frac{\omega}{A}=\frac{n\omega }{g+\frac{b}{b+1}} \label{eq:oscil_phase}
\end{align}

Eq. \ref{eq:oscil_amplitude} reveals that the coupled magnetic and mechanical constraints act as a low-band filter, with a cut-off frequency $\omega_c$:
\begin{equation}
\omega_c = \frac{g + \frac{b}{b+1}}{n}\label{eq:oscil_omegac}
\end{equation}
Considering the values provided in Table \ref{tab:param}, $\omega_c$ ranges between $2$ and $7\cdot 10^4\, s^{-1}$, which, in terms of frequencies, spans the interval [13 Hz -- 450 kHz].
The amplitude of the oscillations is thus modulated by the interplay of the elastic and magnetic contributions to movement, $g$ and $\frac{b}{b+1} = \frac{2 K_D}{2 K_D + m_{sat}B_0}$, and the relative friction torque, $n\omega$. On the other hand, the phase shift $\Phi$ increases with frequency (Eq. \ref{eq:oscil_phase}). Its magnitude is limited by the restoring elastic and magnetic torques that limit the amplitude of the oscillation. In brief, when the friction is dominant (very viscous media or large frequency), the oscillation is damped and in phase quadrature: $\alpha_0 \simeq \theta_0\frac{b}{n\omega(b+1)}\rightarrow 0$ and $\Phi \rightarrow \pi/2$. In the opposite situation, when resistive conservative torques are dominant (predominant elastic material, large magnetic shape anisotropy or low frequency), the amplitude of the oscillation is given by the balance of the magnetic actuation and the elastic and shape anisotropy resistances, and the oscillation remains in phase with the magnetic field. Similarly, when $g\gg1$, the oscillation is limited by the elastic resistance of the material: $\alpha_0 \sim \theta_0b/g(b+1)$ which tends to zero as $g$ increases while $\Phi$ tends to zero. While when $g\ll1$, the oscillation of the particle is limited by the viscous resistance: $\alpha_0 \sim \theta_0/(1 + \frac{n\omega(b+1)}{b})$ and the $\Phi\sim n\omega(b+1)/b$. 

The transition from close to in-phase to close to quadrature oscillating movement is controlled by the cut-off frequency, $\omega_c/(2 \pi)$. $\omega_c$ depends on the viscoelastic parameters, $n$ and $g$. Consistently, increased viscosity or reduced elasticity decreases the value of $\omega_c$ (Eq. \ref{eq:oscil_omegac}). This leads to conclude that particles stimulated in materials with large viscosity or low elasticity compared to the strength of the magnetic anisotropy experience phase-shifted, damped oscillations at lower forcing frequency than those embedded in stiffer or less viscous materials. 

\subsection{Microdisc motion subjected to a rotating magnetic field}

\begin{figure}[h]
\centering
  \includegraphics[height=8cm]{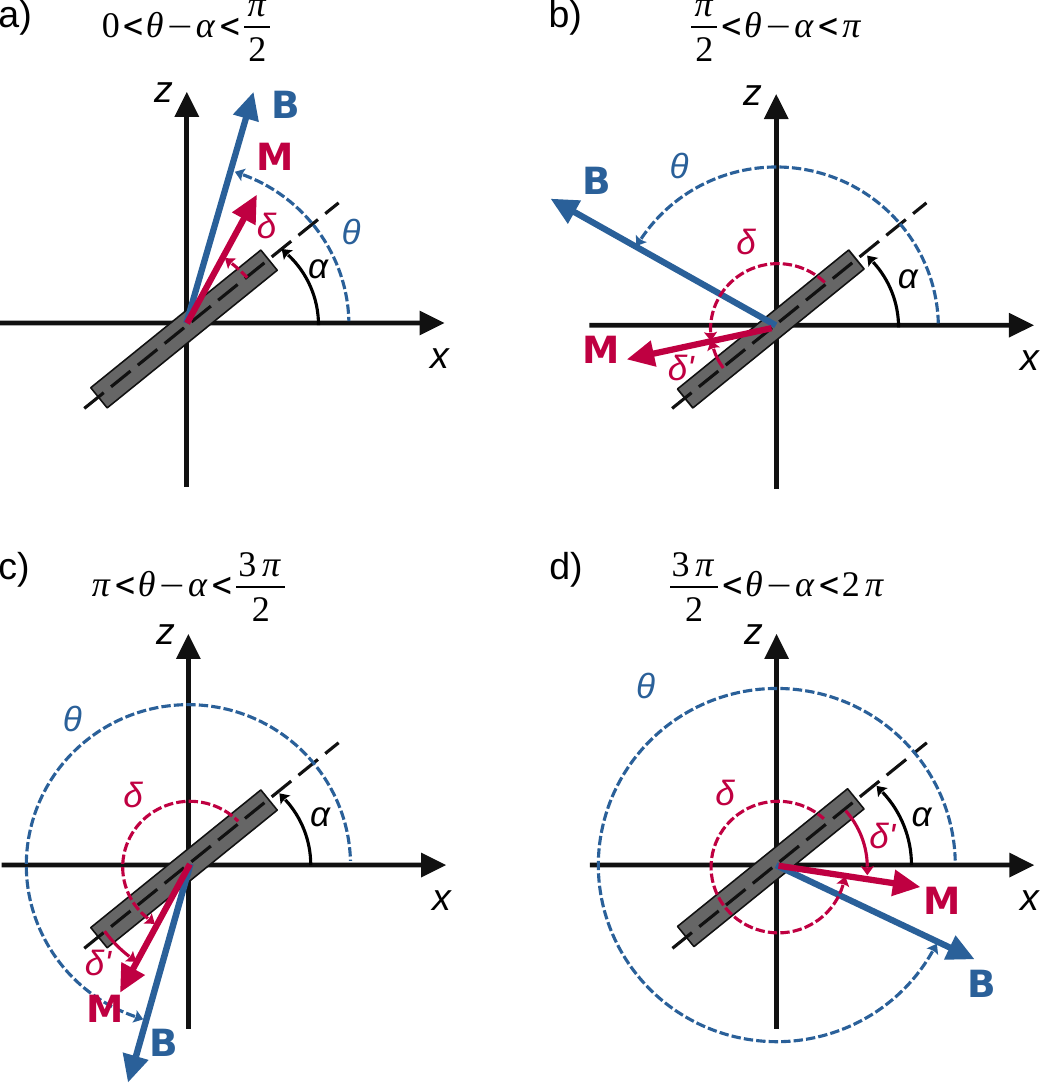}
    \caption{Configurations of the orientation of the magnetisation $\bf{M}$ in relation to the field orientation. Flipping events have occurred between (a) and (b), and (c) and (d).}
    \label{fig:flip}
\end{figure}

We now focus on the scenario where the field rotates continuously in time: $\theta (t) = \omega t$. In this case, the flipping of the magnetisation may become a frequent event depending on the strength of the torques that resist the movement. Magnetisation flip occurs when the angle between the field and the easy plane of the microdisc exceeds $\pi/2$. For a rotating particle, this angle is $\theta-\alpha$ (Fig. \ref{fig:fig4}). When the latter exceeds $\pi/2$ (modulo $\pi$), the magnetisation flips to its symmetric orientation in the particle plane (Fig. \ref{fig:flip}). A similar argument as that given in the previous section shows that the equilibration of the magnetic torques remains governed by the same equation once the magnetisation has flipped (Eq. \ref{eq:proof_flip}). When addressing particle motion, the angle $\theta$ in Eq. \ref{eq:proof_flip} has nevertheless to be changed to $\theta-\alpha$. Just after the flip, the Zeeman torque changes sign, so that it induces an opposite motion that tends to decrease the angle $\alpha$. When the field goes on rotating, the angle between the field and the magnetisation changes sign again which again tends to increase $\alpha$. The magnetisation may thus exhibit four different orientations relative to the field, as illustrated in Fig.  \ref{fig:flip}.

In the following, we analyse three different rheological regimes: a purely viscous fluid (limit case where elasticity is negligible), large values of elastic resistance (limit case where the viscosity is negligible) and an intermediate case where both elasticity and viscosity have to be accounted. The latter is approached qualitatively, as explained below.  

\subsubsection{Motion of the microdisc in a viscous environment}

\begin{figure}[h]
\centering
  \includegraphics[height=9cm]{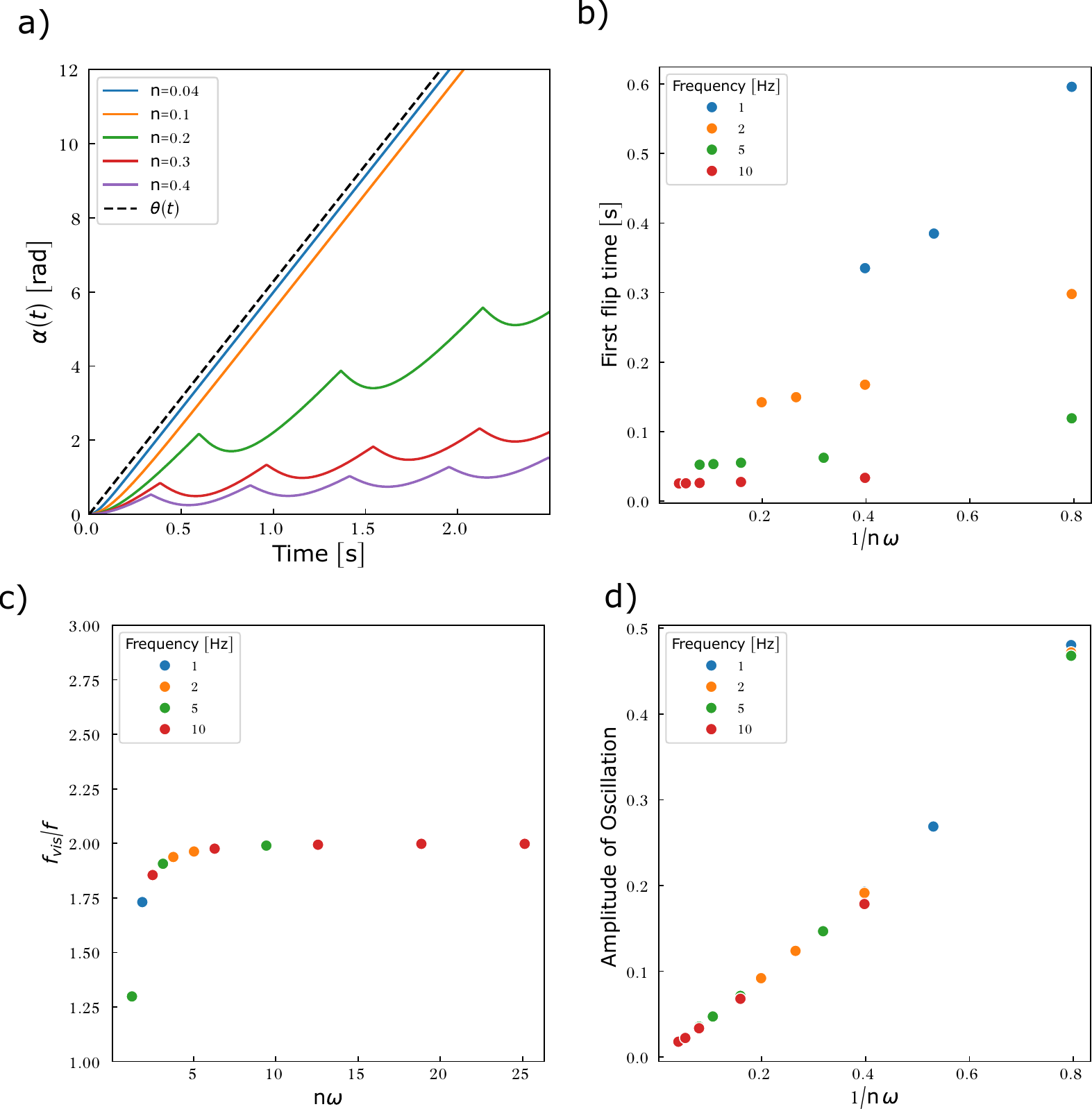}
    \caption{Particle embedded into a predominantly viscous material. a) Particle rotation $\alpha(t)$ compared to the rotation of the magnetic field $\theta(t)$ as function of the reduced viscosity $n$. Frequency = 1 Hz. For values of $n< 0.01$, $\alpha(t)$ cannot be distinguished from $\theta(t)$ (data not shown). b) First flipping time as a function of $1/n\omega$. The points do not gather in a master curve, implying that $n$ and $\omega$ contribute as independent variable to the flipping time.  c) Average frequency of the oscillation $f_{vis}$ as function of $n\omega$. d) Amplitude of oscillation as function of 1/$n\omega$.}
    \label{fig:rotating_vis}
\end{figure}

Eqs \ref{eq:mag}--\ref{eq:mec} are solved numerically in the case where $g=0$ (Fig. \ref{fig:rotating_vis}). As no assumption on $\alpha$ was required to determine the viscous torque, we release the constraint $\alpha \ll 1$ imposed by the calculation of the elastic torque.  This allows the entire viscosity range to be spanned without restriction.

Two regimes of motion are observed, either synchronous or asynchronous with the magnetic field (Fig. \ref{fig:rotating_vis}a). In the synchronous regime, the particle follows the rotating magnetic field with a progressive phase lag that reaches a steady value after a transient regime. This regime is observed at low viscosity. There is no flipping events in this regime. The asynchronous regime arises for larger viscosity. In this regime, the rotational motion of the particle unhinges from the magnetic field, leading to an oscillatory motion. The reason are the flipping events. Because of the friction, $\alpha$ increases less rapidly than $\theta$. At some point, $\theta-\alpha$ reaches $\pi/2$, triggering a flip of the magnetisation. Once the magnetisation has flipped, the torque associated to the Zeeman energy (which has now flipped sign and direction, see Fig.\ref{fig:flip}b) progressively reduces as the field further rotates and the angle between magnetisation and field decreases. Consequently, the energy cost associated to the magnetic shape anisotropy decreases, as the magnetisation falls back to the particle plane, and the mechanical torque coming from the equilibration of the magnetic torques decreases as well. As a result, $\alpha$ decreases until the magnetisation realigns with the field. Once alignment is re-established, the field resumes driving the motion, and $\alpha$ increases again (Fig.\ref{fig:flip}c). The dynamics of the rotation of the particle and of the magnetisation is shown in Supplementary Movie 1. 

We can obtain a crude estimation of the threshold viscosity that leads to flipping events. To this end, we solve analytically Eq. \ref{eq:delta} by setting $g=0$:
\begin{equation}
    \dot{\alpha}(t) = \frac{1}{n}\frac{b\sin(\omega t-\alpha)}{b+\cos(\omega t -\alpha)}
    \label{eq:rot_viscous}
\end{equation}

When the magnetisation is close to flip, the angle $\theta - \alpha$ is close to $\pi/2$. We thus expanded Eq. \ref{eq:rot_viscous} for $\theta-\alpha\simeq\pi/2$ to first order and solved the resulting differential equation. Considering only the steady state regime, we find:
\begin{equation}
\theta - \alpha \simeq \frac{\pi}{2} - (1-n\omega)b
\label{eq:rot_flip_viscous}
\end{equation}

Eq. \ref{eq:rot_flip_viscous} shows that the magnetisation flips as soon as $n\omega \geq 1$. Considering relevant orders of magnitude for $n$, we therefore conclude that the rotation of the microdisc may be transformed into an oscillatory movement when the frequency is in the range $[0.4 - 4000]$ Hz, the lower frequency being attained when the particle is within more viscous media. Since the analytical approach can only be used limitedly, we solved Eq. \ref{eq:rot_viscous} numerically for different parameters as reported in table \ref{tab:param} (Fig. \ref{fig:rotating_vis}).

Consistently, Fig. \ref{fig:rotating_vis}a shows that the particle movement transitions from a uniform to an oscillatory motion once either the forcing frequency or the viscosity exceed a threshold value. In the context of the numerical values used in Fig. \ref{fig:rotating_vis}, the analytical approach suggests that the oscillatory mode should arise as soon as $n \geq 0.16$. Oscillations however appear after a transient time, associated to the first flip of the magnetisation. The dependency of this first flipping time is shown in Fig. \ref{fig:rotating_vis}b as a function of $1/n\omega$. The data do not follow a master curve, meaning that $n$ and $\omega$ contribute as independent variables to this dynamics. Nonetheless as a general result, larger viscosity or larger frequency reduce the time required for the first flipping event to happen. This result is expected as when the viscous torque is dominant, the oscillations are fully dampened ($\alpha(t) \simeq 0$) and the magnetisation flips as soon as $\theta(t) = \pi/2$ (modulo $\pi$). The first flip then occurs at $t=1/(4f)$. This is indeed what is observed in Fig. \ref{fig:rotating_vis}b when $n\omega$ is larger than 5.
We then calculated the mean frequency ($f_{vis}$) and the amplitude of the oscillations of $\alpha(t)$ by numerically extracting local maxima and minima and averaging over the evaluation time window (5 sec). The values are plotted in Fig. \ref{fig:rotating_vis}c,d respectively. As $n\omega$ increases, the particle oscillates at an increasing frequency, that reaches twice that of the forcing field for highly viscous media or at large forcing frequency, as expected. Finally, Fig. \ref{fig:rotating_vis}d shows that the microdisc oscillates with an amplitude governed by the parameter $1/n\omega$ in the range of forcing frequencies studied. 

\subsubsection{Motion of a microdisc trapped in a predominantly elastic material}

\begin{figure}[h]
\centering
  \includegraphics[height=9cm]{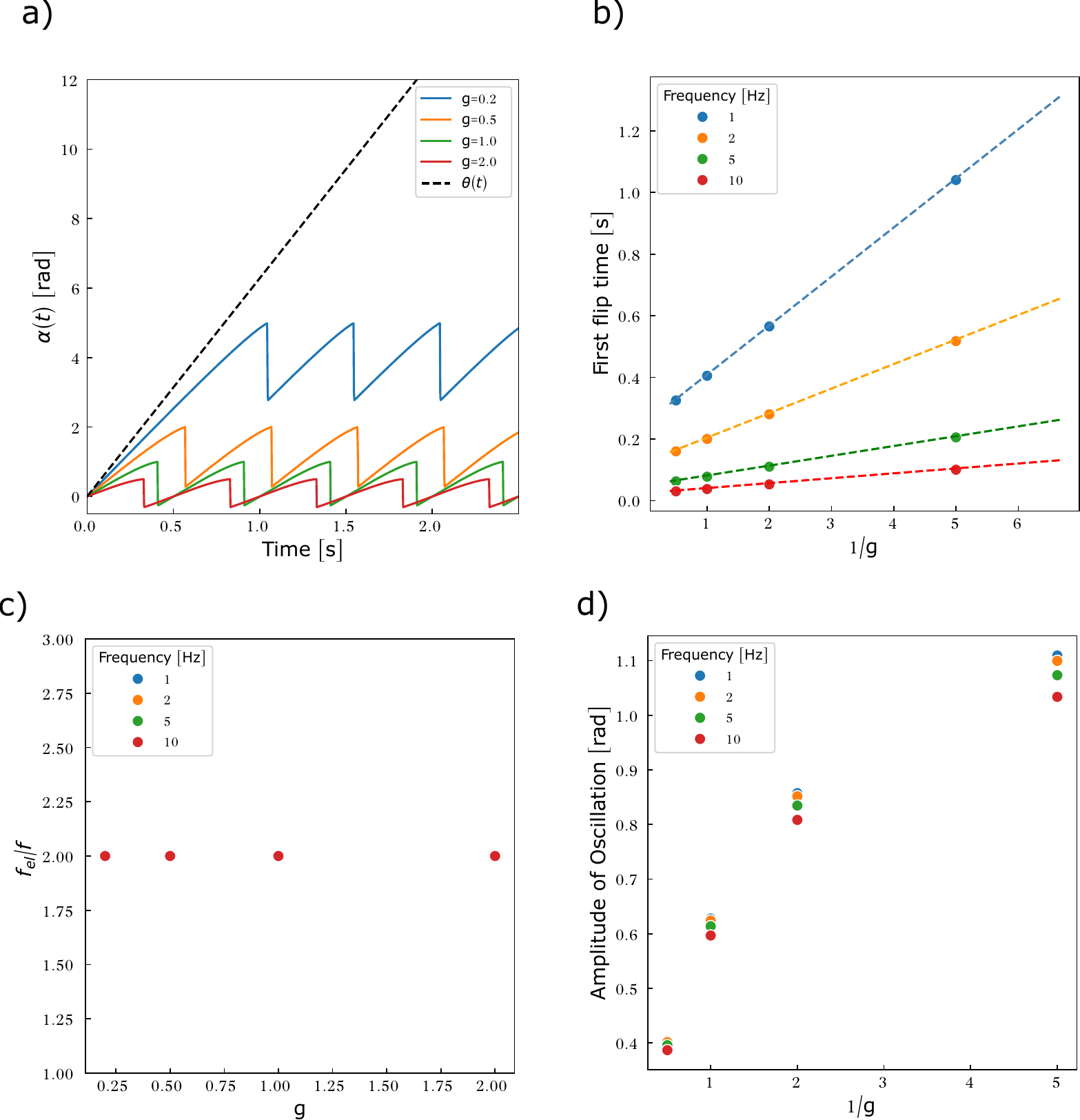}
    \caption{Particle embedded into a predominantly elastic material. a) Particle rotation $\alpha(t)$ compared to the rotation of the magnetic field $\theta(t)$ in relation to the reduced elasticity $g$. Frequency = 1 Hz, $n = 4\times10^{-4}$ s. b) First flipping time as a function of $1/g$ ($\circ$) and analytical expression of the flip time $t_1$ at different frequencies (Eq. \ref{eq:t_el}). c) Average frequency of the oscillation $f_{el}$ as function of $g$. All points overlap at $2f$. d) Amplitude of oscillation as function of 1/$g$. }
    \label{fig:rotation_elast}
\end{figure}

We now analyse the movement of the particle embedded into a predominantly elastic material, exhibiting linear elastic behaviour. $n$ is set to 4$\times10^{-4}$ s in Eqs \ref{eq:mag}--\ref{eq:mec} so that the ratio g/$n\omega$ is [50 to 50000]. In this regime, the elastic contribution is the main resistance to the rotation of the particle. For large enough elastic resistance, $\theta - \alpha$ may reach $\pi/2$, leading to a flipping event. Once the magnetisation has flipped, the magnetic torque favours a backward motion as explained above. As a consequence, at the moment of the flip, a sudden backward jump of the particle occurs due to the instantaneous nature of the elastic stress (Fig. \ref{fig:rotation_elast} and Supplementary Movie 2). 

The backward motion goes in the same direction as the magnetic torque. The subsequent increase in $\alpha$ takes place when the balance between magnetic and elastic torques once again allows the field to drive the particle in the same direction as the field. When the coupling parameter $g$ is low, the particle oscillates around a mechanically stressed configuration (Fig. \ref{fig:rotation_elast}a, blue or yellow curves). Conversely, for large values of $g$,  the particle oscillates around its resting position (Fig. \ref{eq:max_rot}a, red curve). This latter case arises because the particle moves only slightly away from its equilibrium before the magnetisation flips, and the elastic stress fully relaxes. The flipped magnetisation then experiences an opposite torque that leads to negative values of $\alpha$, until the field realigns with the particle plane and resumes driving the motion (see Supplementary Movie 2). Note that the blue curve of Fig. \ref{fig:rotation_elast}a ($\alpha(t)$ for $g=0.2$) is way beyond the approximation $\alpha\ll1$ and should be taken more as a qualitative description.   

The oscillatory motion of the particle can be approached analytically when the angle between the magnetisation and the easy plane of the particle is small. Before the first flip of the magnetisation, $\delta \ll1$, and the motion is governed by Eq. \ref{eq:delta} with $n\simeq0$. Just before the magnetisation flips, the angle between the field and the magnetisation is close to $\pi/2$. As in the viscous case, we  solve Eq. \ref{eq:delta} for $\theta - \alpha \simeq \pi/2$. We obtain that the first magnetisation flip occurs at time $t_1$:

\begin{equation}
t_1 \simeq \frac{1}{g\omega} + \frac{\pi}{2\omega} \label{eq:t_el}
\end{equation}

Figure \ref{fig:rotation_elast}b shows the evolution of the first flipping time with the stiffness of the surrounding material, obtained from the numerical resolution of Eq. \ref{eq:delta}. Consistent with the analytical solution Eq. \ref{eq:t_el}, we find that the first flipping time is inversely proportional to $g$. The dashed lines in Fig. \ref{fig:rotation_elast}b represent the Eq. \ref{eq:t_el} for different $\omega$. The second flip of the magnetisation occurs when $\theta-\alpha \simeq 3\pi/2$. The motion is governed by Eqs \ref{eq:mag}--\ref{eq:mec}  in this regime, except that the small angle is no more $\delta$ but $\delta' = \delta - \pi$ (Fig. \ref{fig:flip}c). Expanding Eq. \ref{eq:mag}--\ref{eq:mec} to first order in $\delta'\ll 1$, we find the second flipping time: 

\begin{equation*}
t_2 \simeq \frac{1}{g\omega}+\frac{3\pi}{2\omega} 
\end{equation*}

This calculation can be made general for any flip of the magnetisation. We therefore conclude that the magnetisation flips at a frequency $f_{el}$:
\begin{equation}
f_{el} = 2 f
\label{eq:rot_elast_f}
\end{equation}

that is independent of the stiffness of the material. The ratio $f_{el}/f$ obtained from the numerical calculation of $\alpha$ is shown in Fig. \ref{fig:rotation_elast}c. Consistently we observe that all data stack at 2, whatever the stiffness of the outer material. The elastic coefficient of the material only enters the expression of the transient duration before the magnetisation starts flipping. Expectedly, it flips sooner in stiffer materials than in softer ones as particle rotation is more impeded. 
Lastly we observe that the amplitude of the oscillation decreases as the material becomes stiffer (Fig. \ref{fig:rotation_elast}d). A small dependence on the forcing frequency is visible. The cause is the small contribution of $n$ that gains importance at low values of $g$. As previously discussed, the amplitude of oscillations are attenuated by the viscous torque in a frequency dependent manner.

\subsubsection{Motion of a microdisc in a viscoelastic material, the viscous friction being comparable to the elastic resistance}

Finally we propose a qualitative description in the case where the elastic resistance is comparable to the viscous one. Indeed, in this scenario $\alpha\ll1$ is no longer valid and our estimation of the elastic torque in principle fails. Indeed, we already inaccurately explored regimes of large $\alpha$ in the previous section (see Fig. \ref{fig:rotation_elast}a). Large deformations are expected to distribute the elastic energy afar from the moving particle. The energy is restored non locally, therefore making the material appears stiffer than it is \cite{chadwick_oscillations_1966, chadwick_torsional_1969}. As a consequence, the oscillations are expected to be more damped than predicted by our first-order modelling. By computing  Eq. \ref{eq:mag}--\ref{eq:mec}, our goal here is to get an intuition of how the oscillating movement of the particle is modified when both viscosity and elasticity impair the movement.

\begin{figure}[h]
\centering
  \includegraphics[height=9cm]{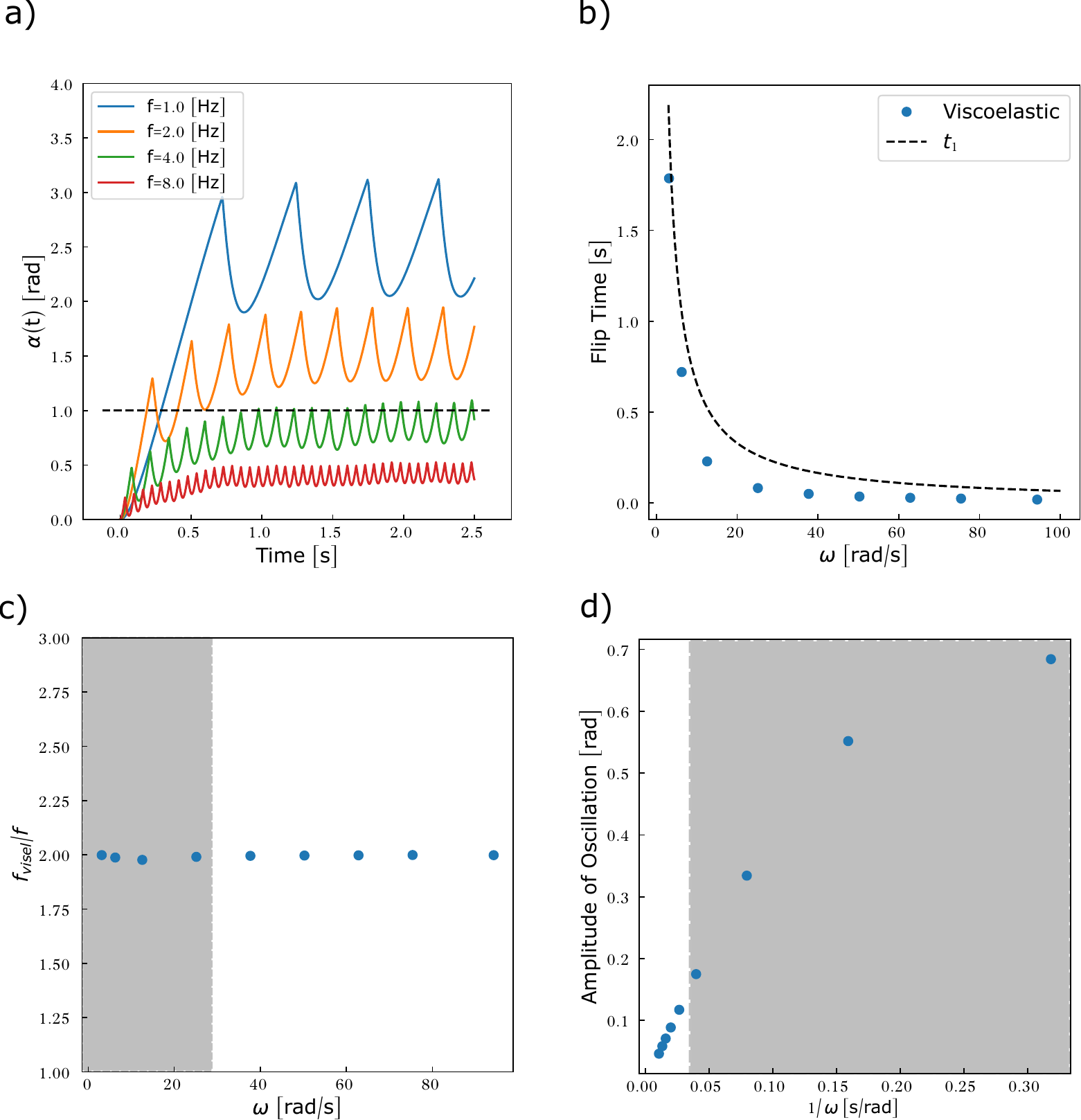}
    \caption{Particle embedded into a visco-elastic material. a) Particle rotation $\alpha(t)$ for different field rotating frequencies $f$. The dashed line indicates the limit of the validity of $\alpha\ll1$. $n=0.1$ s, $g=0.2$. b) First flipping time as a function of $\omega$. The expression of analytical derived $t_1$ (Eq. \ref{eq:t_el}) is plotted as a dashed black line.  c) Average frequency of the oscillation $f_{vis}$ as function of $\omega$. d) Amplitude of oscillation as function of $\omega$. The grey rectangle indicates the region in which the approximation $\alpha\ll1$ is valid.}
    \label{fig:rotation_viscoelas}
\end{figure}

Our first observation is that the oscillatory motion is now a complex combination of what we reported in the cases of predominantly viscous and predominantly elastic material (Fig. \ref{fig:rotation_viscoelas}a and Supplementary Movie 3). The particle oscillates up to a plateau amplitude, that originates from the elastic resistance. Keep in mind that this maximal deformation is expected to be overestimated. When the magnetisation flips, the elastic stress is relaxed within a time that depends on the viscosity. This time is not solely the Maxwell relaxation time but involves the magnetic shape  anisotropy as it was the case in a predominantly viscous medium. Fig. \ref{fig:rotation_viscoelas}b shows the computed first flipping time, compared to the one expected in a predominantly elastic environment (Eq. \ref{eq:t_el}). The discrepancy between the analytical approach and the numerical solution highlights that viscosity and elasticity contribute in an indissociable manner to the dynamics of the flipping events. 

Fig. \ref{fig:rotation_viscoelas}c illustrates the complexity brought by the combined viscous and elastic resistances to the magnetic actuation. The frequency of the oscillations of the particle now varies non monotonically with the forcing frequency. Indeed, at low forcing frequency, the dominant resistance to the particle motion comes from the elastic properties of the material. Therefore the particle oscillates at a frequency very close to the double frequency, as expected from Eq. \ref{eq:rot_elast_f}. At large frequency, both the viscous and the elastic resistance of the material force the oscillation to take place at the double frequency. But at intermediate frequency, when the viscous torque is intermediate but comparable to the elastic torque, the frequency of oscillation is reduced, as it was observed in a viscous medium (Fig. \ref{fig:rotating_vis}c). Finally, Fig. \ref{fig:rotation_viscoelas}d highlights the impact of the forcing frequency on the amplitude of the motion in a viscoelastic environment, supporting the hypothesis made in the previous section on the origin of the dispersion of data obtained in Fig. \ref{fig:rotation_elast}d).  

\section{Discussion}

In this study, we modelled the magnetically induced motion of a micrometric particle embedded in a biological viscoelastic material, actuated by either oscillating or rotating magnetic fields. This configuration has been widely employed in experimental studies, both to probe the rheological properties of biological materials and as a proof of concept for innovative therapeutic strategies \cite{naud_cancer_2020, dieny_magnetism_2025}. Despite the growing body of experimental work in this domain, the underlying physical framework is only partially explored. This gap between theory and experiment leads to the absence of clear guidelines when selecting parameters for magnetic particle actuation. The objective of this work was to provide a deeper understanding of the physical interactions underlying magneto-mechanical stimulation, focusing on how key parameters such as field frequency, viscosity, and stiffness influence particle dynamics. Ultimately, the proposed model offers a simplified yet predictive framework to anticipate first-order effects.

The principal finding of this study is that the use of a rotating magnetic field typically leads to oscillatory particle behaviour across most of the scenarios explored. In particular, we found that in a predominantly viscous material, as soon as the condition $n = 1/\omega$ is met, the particle motion transitions into an asynchronous regime. This critical frequency threshold has also been reported by Berret et al. \cite{berret_local_2016}, where it was used to infer the viscoelastic properties of the cell interior. Experimentally observed values for the critical frequency $\omega_c$ are on the order of 0.1 rad·s$^{-1}$ ($\sim$0.2 Hz), corresponding to materials with viscosities in the tens of Pa·s and stiffnesses around 10 Pa—parameters that fall at the lower end of the range explored in our study. As a result, the synchronous regime is only observed under conditions of low viscosity, low stiffness, and low actuation frequency. From a practical standpoint, this result offers useful flexibility to experimentalists: when operating in conditions that favour the asynchronous regime, rotating and oscillating magnetic fields can be used interchangeably. Notably, oscillating fields offer the additional advantage of enabling direct control over the asymptotic amplitude via the parameter $\theta_0$. 

Focusing on the oscillatory motion, regardless of the origin field, we estimated that the amplitude of oscillation, $\Delta\alpha$, is between 0.2-0.5 rad (for intermediate viscoelastic parameters). This quantity can be converted with a simple calculation into a displacement $U=\Delta\alpha\times R$, with $R$ the radius of the particle. In our case this leads to displacements of the order 0.1-0.3 $\mu$m at the very edge of the particle. Typical length of collagen 1 and actin fibres are in the $\mu$m range. Therefore the particle motion can lead to important deformation of networks composed of such proteins. The amplitude of the oscillation was shown to be sensitive to two adimensional parameters $g$ and $n\omega$ which represent the competition of elasticity and viscosity to the magnetic field respectively (Eq. \ref{eq:reduced}). As both are increased, our model predicts that the amplitude of oscillation is attenuated. Theoretical and experimental evidence of similar trends have been reported either in theoretical approaches or experimental studies. For instance, Wilhem et al. \cite{wilhelm_rotational_2003} modelled the actuation of magnetic chains composed of paramagnetic beads with oscillating fields in a Maxwellian fluid and showed that the amplitude of oscillation of rod-like particle would decreases as the forcing frequency increases. From an experimental point of view, Kim et al. \cite{kim_biofunctionalized_2010} used micrometric magnetic discs to stretch the membrane of glioma cells and activate mechano-responsive pathways. Their data show that the efficiency of the stimulation decreases as the field frequency goes beyond 20 Hz.

It is to be kept in mind that the modelling we propose is a 2D description. This simplified view was proposed to enable a more accessible analysis, supported by the analytical description of limit cases. It has nevertheless a critical impact as it makes the model “forget” that the microdisc can rotate and align its easy magnetic plane with the plane of rotation of the field. And in that case, the magnetic torque does not induce any mechanical movement since the magnetisation rotates in the easy-plane of the particle \cite{mansell_magnetic_2017}. This limitation is inherent to using a 2D framework to describe vortex particles. For instance it was not present in the work of Berret et al. \cite{berret_local_2016}, where rod-shaped particles were studied. Then our description is expected to fail when both the elastic and the viscous torques are so low compared to the Zeeman torque that the realignment of the microdisc plane with the plane in which the field rotates occurs in a time that is short compared to the duration of the experiment. So far, the actuation of vortex microdiscs in biological environments has always shown significant effects on cells, probably because the particles are not dispersed into fluid-like compartments but are indeed integrated into macromolecules networks or interacting with membranes that offer a significant viscoelastic resistance. A 3D numerical analysis would however be necessary to describe entirely this mechano-magnetic coupling. This study goes beyond our objective, which was to obtain initial insights into the impact of viscoelastic environments on the movement of magnetically actuated vortex microdiscs, relevant for biological applications. It will however be of interest to address it in a near future.

 The linear rheological description we used is also to be challenged. Maxwell’s model is commonly used as a first-step approach for probing the behaviour of complex viscoelastic systems such as cell cytoplasm \cite{berret_local_2016} or micellar solutions \cite{wilhelm_rotational_2003}. This choice inherently assumes first, a continuum material description, and second, small deformations of the elastic components. The validity of this assumption is assessed by comparing the size of the microparticles with the typical mesh size of the biological networks considered. As discussed in the theoretical section, the porosity and mesh size of such biomaterials depend on factors such as concentration and polymerisation conditions. For example Matrigel exhibits a pore size between 100-200 nm \cite{zaman_migration_2006} while the pore size in gels made of collagen 1 varies between 1 to 10 micrometre, values that are thus comparable to the dimensions of the particles used in this study \cite{roeder_tensile_2002}. When addressing cell interior, the microdisc may either be trapped in the cytoskeleton of the cell or embedded in the cytoplasm. The mesh size of the actin cortex is about 50 nm  \cite{jonathon_howard_mechanics_2001}, far less than the size of the particles. For similar size-related reason, the cytoplasm can be modelled as a viscoelastic material in which the particles feel the elasticity coming from the presence of various organelles inside it and the intermediate filaments \cite{arjona_cytoplasm_2023}. The choice of using a Maxwell model and not to limit to a fluid description comes from the assumption that, since the frequencies assayed are in the Hz range, we expect that vibrating particles remain trapped into the mesh of proteins and do not repel these viscoelastic structures far from them. Nevertheless, in the rotating regime, deformations can exceed the linear regime and a reorganisation of the biological compounds could take place, leading to plastic behaviour.

In the presence study, we limit our description of the elastic resistance to first order. However, the key dependencies of the elastic coupling assumed in our model are consistent with findings from previous studies. For instance, a similar formulation of the torsional constant was employed by Wilhelm et al. \cite{wilhelm_rotational_2003}, where the torque damping coefficient is given by $\gamma = \kappa_e V G$, with $\kappa_e$ being a dimensionless geometric factor, $V$ the particle volume (scaling with the cube of a characteristic dimension), and $G$ the shear modulus, itself proportional to the Young's modulus $E$. Similarly, Berret et al. \cite{berret_local_2016} modelled the elastic resistance of a microrod as a product of a shape-dependent factor, the shear modulus $G$, and the cube of the rod length ($L^3$). These examples illustrate that by appropriately tuning the geometric prefactor in the expression for $\gamma$, our modelling framework can be adapted to different particle geometries.

Actuation of superparamagnetic iron oxide nanoparticles (SPIONs) have been used for long to induce thermal effects in biological samples or tissues \cite{gavilan_magnetic_2021}. The heat source is either magnetic or mechanical, the latter referred to as Brownian dissipation. For small particles like SPIONs the Brownian dissipation is often negligible and the thermal effect originates from magnetic losses following high frequency actuation \cite{pucci_superparamagnetic_2022}. However for large anisotropic particles, the mechanical contribution could become important. Indeed, thermal energy is released in vicinity of the moving microdisc, coming from the viscous friction of the particle with the outer medium. 
To evaluate thermal losses associated to viscous friction, we considered microdiscs embedded into a purely viscous medium and use a scaling law approach to calculate the energy loss following their actuation:

\begin{equation}
    P_{vis} = \frac{dE_{vis}}{dt} = \nu \ddot{\alpha} \simeq \eta R^3 \Delta\alpha f^2
\end{equation}

This energy increases with frequency while it is accompanied by a decreased displacement. For intermediate values of $\eta$ and using values for $\Delta \alpha$ coming from our numerical calculation, we obtain that the dissipated power is of the order of few fW (2 fW for a frequency of 10 Hz). This quantity is to be compared to the thousands W/g that are at play in magnetic heat generation for hyperthermia \cite{liu_comprehensive_2020}. Using the weight of a standard permalloy microdisc ($\sim1\times10^{-12}$ g), we obtain that viscous friction generate energy transfer of the order of 0.002 W/g, which is several order of magnitude smaller than the targeted values for causing cellular dysfunctions. The significant difference in the orders of magnitude clearly separates mechano-stimulation from hyperthermia. 

The trend is different when the dominant term is elasticity. Such energy is stored in the surrounding environment and it can be estimated as follows in the proximity of the particle edge: 

\begin{equation}
   E_{el} \simeq \gamma \Delta\alpha^2 = ER^3\Delta\alpha^2
\end{equation} 

By limiting our analysis to the particle edge, we avoid the limitation of our elastic description that does not take into account stress propagation far from the particle. This stored energy increases with the stiffness of the outer material, while the amplitude decreases and is independent of the frequency. For intermediate values for $E$, using values from our numerical calculation, we obtain that the mechanical energy is of the order of 0.1 fJ. Converted in $k_B$T units, this amounts to about $2\times10^5$ $k_B$T at 37 \textcelsius. Cells spend 30.5 kJ/mol to phosphorylate ATP into ADP, which corresponds to about 10 $k_B$T per molecule at 37 \textcelsius. Thus the motion of a particle could ideally transfer an energy that corresponds to two thousands ATP molecules. To get a clearer idea of how large this value is, one can compare to a common energy-consuming cellular process such as actin treadmilling \cite{alberts_molecular_2022}. For treadmilling to take place, a critical concentration of actin-ATP of 0.16 $\mu$M is required \cite{boal_mechanics_2012}. By taking the volume of one single vortex particle as reference, this value corresponds to about 6 molecules of ATP per particle, which expressed in energy terms, is 60 $k_B$T. This suggests that the mechanical actuation of one single particle transmits much more energy to the cell than what is involved in the polymerisation of the cytoskeleton. Our calculation therefore leads to conclude that cellular alterations following magnetic stimulation with such particles are to be attributed to mechanical origin and not to thermal dissipation, in consistence with the conclusion reached experimentally in Kim et al. \cite{kim_biofunctionalized_2010}.

\section{Conclusions}
In conclusion, this study introduced a 2D model of magnetic vortex microdisc dynamics in viscoelastic environments under oscillating and rotating magnetic fields. It showed how rheological properties and field frequency control particle motion, including the transition between synchronous and asynchronous regimes. The energy associated with magnetically induced motion was found to generate negligible heat while producing mechanical stresses that can, in principle, compete with biomolecular forces. While the simplified approach captures key physical behaviours, extending it to 3D will be essential to fully describe disc reorientation over time, particularly under low viscoelastic resistance. Despite these limitations, the model provides a practical tool to anticipate the influence of viscosity, stiffness, and frequency on vortex microdisc actuation, offering valuable guidance for the design of magneto-mechanical stimulation experiments and their biomedical applications.

%%%%%

\section*{Author Contributions}
AV, AN, and BD designed the model. AV performed the analytical and numerical calculations under the supervision of RM, HJ, BD, and AN. HJ, BD, RM, and AN secured funding for the project. AV and AN wrote the first draft, which was reviewed and edited by all authors.

\section*{Conflicts of interest}
There are no conflicts to declare.

\section*{Acknowledgements}
This work is supported by the French National Research Agency in the framework of the “Investissements d'avenir” program (ANR-15-IDEX-02).

\printbibliography

\newpage
\begin{center}
{\huge \bf Supplementary information}\\
\vspace{0.5cm}
{\LARGE Modelling of magnetic vortex microdisc dynamics under varying magnetic field in biological viscoelastic environments}\\
\vspace{0.5cm}
Andrea Visonà$^{1,2}$, Robert Morel$^{2}$, Hélène Joisten$^{2}$, Bernard Dieny$^{2}$, and Alice Nicolas$^{1}$\\
$^1$ Univ. Grenoble Alpes, CNRS, CEA/LETI-Minatec, Grenoble INP, LTM, Grenoble F-38000, France.\\
$^2$ Univ. Grenoble Alpes, CEA, CNRS, Spintec, Grenoble F-38000, France.
\end{center}
\setcounter{section}{0}
\renewcommand{\thesection}{S\arabic{section}}
\section{Derivation of the demagnetising energy term}

Writing the second Maxwell equation in the absence ocellular alterations following stimulation with such particles are to be attributed to mechanical origin external field yields $\nabla \cdot B = \mu_0 \nabla \cdot (H_d + M) = 0$  and hence $\nabla H_ d = - \nabla \cdot M$, with $B$ the magnetic induction, $H_d$ the demagnetising field, $M$ the magnetisation and $\mu_0$ the free space permeability. Note that here, only the part of the dipolar field occurring inside the magnetic body, the so-called demagnetising field, is considered. From this equation, it is possible to define magnetic volume charges as $\rho_m = - \nabla \cdot M$ in the bulk and surface charges $\sigma_m = \textbf{M}\cdot \hat{n}$ where $\hat{n}$ is the normal to the surface. When considering only the dipolar interaction, the ground state is then given by the magnetic configuration that minimises the volume and magnetic surface charges, that will minimise the demagnetising field and therefore minimises the energy. Thus, the shape of the sample strongly influences the demagnetising energy, giving rise to an anisotropy referred to as shape anisotropy and some preferential axis or planes where the magnetisation prefers to lay. 
To illustrate this, let us consider the case of a uniformly magnetised thin ellipsoid with principal axis along ($x,y,z$) such that the semiaxis R $>>$ h. 

The demagnetising field can be expressed simply in terms of the demagnetising coefficients $N_i$ as: 
\begin{equation}
   H_{d,i} = - N_iM_i 
\end{equation}
In the chosen geometry the tensor N has non-zero terms only in the diagonal and its trace is equal to 1. The demagnetising field in polar spherical coordinates ($\theta, \phi$), reads as:

\begin{equation}
   H_d = -N \textbf{M} = 
  - \left( \begin{array}{ccc} N_R & 0 & 0 \\  0 & N_R & 0 \\   0 & 0 & N_h \end{array} \right) M_{sat}\left( \begin{array}{c} \cos\phi\sin\theta \\ \sin\phi\sin\theta \\ \cos\theta \end{array} \right) = 
  - M_{sat}\left( \begin{array}{c} N_R\cos\phi\sin\theta \\ N_R\sin\phi\sin\theta \\ N_h\cos\theta \end{array} \right)
\end{equation}

The energy is calculated as:

\begin{equation}
   \varepsilon_d = -\frac{1}{2}  \mu_0 \textbf{M}\cdot\textbf{H}_d = \frac{1}{2}\mu_0M_{sat}^2 \left( \begin{array}{c} \cos\phi\sin\theta \\ \sin\phi\sin\theta \\ \cos\theta \end{array} \right) \left( \begin{array}{c} N_R\cos\phi\sin\theta \\ N_R\sin\phi\sin\theta \\ N_h\cos\theta \end{array} \right) 
\end{equation}

which becomes: 

\begin{equation}
\varepsilon_d = \frac{1}{2}\mu_0M_{sat}^2 \left( N_R \sin^2\theta + N_h \cos^2\theta\right) 
\end{equation}

with the relation $N_R = \frac{1 - N_h}{2}$

We now introduce a more meaningful angle, $\delta$, which is the angle between the easy plane of the ellipsoid and the magnetisation vector and rewrite the demagnetisation energy as function of it. Indeed $\delta$ allows to easily indicate the in plane and out of plane component of the magnetisation. Being $\delta = \pi/2-\theta$

\begin{equation}
\varepsilon_d = \frac{1}{2}\mu_0M_{sat}^2 \left( N_R \cos^2\delta + N_h \sin^2\delta\right) 
\end{equation}

Expressing the previous equation as energy ($E$) instead of energy density ($\varepsilon$) we obtain: 

\begin{equation}
    E_d = \frac{1}{2}\mu_{0}\frac{m_{sat}^2}{V}\left( N_R \cos^2\delta + N_h \sin^2\delta\right)
\end{equation}

\section{Supplementary movies}

\begin{itemize}
    \item Motion of the microdisc in a viscous environment: same as Fig. 8 ($n=0.2$, $g=0$), \textbf{Supplementary\_movie\_1.mp4}
    \item Motion of a microdisc trapped in a predominantly elastic material: same as Fig. 9 ($n=4\times10^{-3}$, $g = 2$), \\ \textbf{Supplementary\_movie\_2.mp4}
    \item Motion of a microdisc in a viscoelastic material, the viscous friction being comparable to the elastic resistance: \\ same as Fig. 10 ($n=0.1$, $g = 0.2$), \textbf{Supplementary\_movie\_3.mp4}\\
\end{itemize}
\end{document}